\documentclass[10pt,a4paper,onecolumn]{article}
\usepackage[a4paper, total={6in, 9.8in}]{geometry}
\usepackage{setspace,url}
\usepackage{lineno,hyperref,multirow,cite}
\usepackage{authblk}
\usepackage{epsfig,amssymb,amsmath}
\usepackage[utf8]{inputenc}
\usepackage{amsmath,graphicx}
\usepackage{amssymb,amsmath,epsfig,url,mathrsfs} \usepackage{algorithm,algorithmic}
\usepackage[usenames, dvipsnames]{color}
\usepackage[framemethod=TikZ]{mdframed}
\usepackage{xcolor}
\usepackage{colortbl}
\usepackage[colorinlistoftodos]{todonotes}
\usepackage{multirow}
\usepackage{multicol}
\usepackage{enumitem}

\usepackage{numprint}
\npthousandsep{,}

\usepackage{graphicx}
\usepackage{subfig}
\usepackage{array, makecell}
\usepackage{verbatim}
\usepackage[utf8]{inputenc}
\usepackage[T1]{fontenc}
\usepackage[ngerman,english]{babel}
\usepackage{tabularx}  
\usepackage{ragged2e}  
\newcolumntype{Y}{>{\RaggedRight\arraybackslash}X} 
\usepackage{booktabs}  
\usepackage{amsthm}
 
\usepackage{tcolorbox}
\usepackage{graphicx}
 
\definecolor{chestnut}{rgb}{0.8, 0.36, 0.36}
\definecolor{frenchblue}{rgb}{0.0, 0.45, 0.73}
\definecolor{babyblue}{rgb}{0.54, 0.81, 0.94}
\definecolor{beaublue}{rgb}{0.74, 0.83, 0.9}
\definecolor{bubblegum}{rgb}{0.99, 0.76, 0.8}
\definecolor{gainsboro}{rgb}{0.86, 0.86, 0.86}
\definecolor{amethyst}{rgb}{0.6, 0.4, 0.8}
\definecolor{mygray}{gray}{0.85}
\definecolor{aq}{rgb}{0.5, 1.0, 0.83}
\definecolor{aquamarine}{rgb}{0.5, 1.0, 0.83}
\definecolor{beaublue}{rgb}{0.74, 0.83, 0.9}
\definecolor{cadetblue}{rgb}{0.37, 0.62, 0.63}
\definecolor{lavender}{rgb}{0.9, 0.9, 0.98}
\definecolor{nonphotoblue}{rgb}{0.64, 0.87, 0.93}
\definecolor{paleaqua}{rgb}{0.74, 0.83, 0.9}
\definecolor{peachpuff}{rgb}{1.0, 0.85, 0.73}
\definecolor{uscgold}{rgb}{1.0, 0.8, 0.0}
\definecolor{pastelorange}{rgb}{1.0, 0.7, 0.28}
\definecolor{mountainmeadow}{rgb}{0.19, 0.73, 0.56}
\definecolor{mossgreen}{rgb}{0.68, 0.87, 0.68}
\definecolor{lemonchiffon}{rgb}{1.0, 0.98, 0.8}
\definecolor{lightblue}{rgb}{0.68, 0.85, 0.9}
\definecolor{azuremist}{rgb}{0.94, 1.0, 1.0}
\definecolor{caribbeangreen}{rgb}{0.0, 0.8, 0.6}
\definecolor{celadon}{rgb}{0.67, 0.88, 0.69}
\definecolor{camouflagegreen}{rgb}{0.47, 0.53, 0.42}
\definecolor{coolgrey}{rgb}{0.55, 0.57, 0.67}
\definecolor{darkolivegreen}{rgb}{0.33, 0.42, 0.18}
\definecolor{darkspringgreen}{rgb}{0.09, 0.45, 0.27}

\makeatletter
\newcommand{\mathleft}{\@fleqntrue\@mathmargin0pt}
\newcommand{\mathcenter}{\@fleqnfalse}
\makeatother
 
\theoremstyle{definition}

\usepackage{todonotes}

\newcommand{\quotebox}[1]{\begin{center}\fcolorbox{white}{blue!10!gray!10}{\begin{minipage}{0.99\linewidth}\vspace{3pt}\center\begin{minipage}{0.95\linewidth}{\space\large``}{#1}{\large''}\end{minipage}\smallbreak\end{minipage}}\end{center}}

\hypersetup{
   unicode=false,          
   pdftoolbar=true,        
   pdfmenubar=true,        
   pdffitwindow=false,     
   pdfstartview={FitH},    
   pdftitle={VoicePrivacy 2022 Challenge Evaluation Plan},    
   pdfauthor={VoicePrivacy},     
   pdfsubject={VoicePrivacy},   
   pdfcreator={},   
   pdfproducer={}, 
   pdfkeywords={VoicePrivacy, anonymisation}, 
   pdfnewwindow=true,      
   colorlinks=true,       
   linkcolor=blue,          
   citecolor=blue,        
   filecolor=magenta,      
   urlcolor=blue           
}

\usepackage{color, colortbl}
\definecolor{LightCyan}{rgb}{0.88,1,1}

\mdfdefinestyle{MyFrame}{%
    linecolor=black,
    outerlinewidth=.1pt,
    roundcorner=1pt,
    innertopmargin=\baselineskip,
    innerbottommargin=\baselineskip,
    innerrightmargin=5pt,
    innerleftmargin=-1pt,
    backgroundcolor=gray!10!white}

\title{The VoicePrivacy 2022 Challenge\\ Evaluation Plan\\[1em]\large{}Version 1.0}

\author[1]{Natalia Tomashenko}
\author[2]{Xin Wang}
\author[2]{Xiaoxiao Miao}
\author[3]{Hubert Nourtel}
\author[3,4]{Pierre Champion}
\author[5]{Massimiliano Todisco}
\author[3]{Emmanuel Vincent}
\author[5]{Nicholas Evans}
\author[2,6]{Junichi Yamagishi}
\author[1]{Jean-François Bonastre}

\affil[1]{Laboratoire Informatique d’Avignon (LIA), Avignon Université, France}
\affil[2]{National Institute of Informatics, Tokyo, Japan}
\affil[3]{Université de Lorraine, CNRS, Inria, LORIA, France}
\affil[4]{LIUM, Le Mans Université, France}
\affil[5]{Audio Security and Privacy Group, EURECOM, France}
\affil[6]{University of Edinburgh, UK}

\date{\url{https://voiceprivacychallenge.org}}

\begin{document}

\maketitle

\begin{tcolorbox}[width=\textwidth, colback={azuremist}, title={\textbf{For new participants --- Executive summary}}, colbacktitle=lightblue, coltitle=black, arc=0.3mm, fonttitle=\bfseries, boxrule=0.5pt]    
  \begin{itemize}
      \item The task is to develop a voice anonymization system for speech data which 
      conceals the speaker's voice identity while protecting linguistic content, paralinguistic attributes,
      intelligibility and naturalness.
      \item Training, development and evaluation datasets are provided in addition to 3 different baseline anonymization systems, evaluation scripts, and metrics. Participants apply their developed anonymization systems, run evaluation scripts and submit objective evaluation results and anonymized speech data to the organizers.
      \item Results will be presented at a workshop held in conjunction with  
      INTERSPEECH 2022
      to which all participants are invited to 
      present their challenge systems and to
      submit additional workshop papers.
  \end{itemize}
\end{tcolorbox}

\vspace{5pt}

\begin{tcolorbox}[width=\textwidth, colback={azuremist}, title={\textbf{For readers familiar with the VoicePrivacy Challenge --- Changes w.r.t.\ 2020}}, colbacktitle=lightblue, coltitle=black, arc=0.3mm, boxrule=0.5pt]    
   \begin{itemize}
     \item A stronger, semi-informed \textbf{attack model} in the form of an automatic speaker verification (ASV) system trained on anonymized (per-utterance) speech data.
     \item Complementary \textbf{metrics} comprising the equal error rate (EER) as a privacy metric, the word error rate (WER) as a primary utility metric, and the pitch correlation $\rho^{F_0}$ and gain of voice distinctiveness $G_{\text{VD}}$ as secondary utility metrics.
     \item A new \textbf{ranking} policy based upon a set of minimum target privacy requirements.
   \end{itemize}
\end{tcolorbox}

\newpage

\section{Context}

Recent years have seen mounting calls for the preservation of privacy when treating or storing personal data. This is not least the result of recent European privacy legislation, e.g., the general data protection regulation (GDPR). While there is no legal definition of privacy~\cite{nautsch2019gdpr}, speech data is likely to fall within the scope of privacy regulation.   Speech encapsulates a wealth of personal, private data, e.g., age and gender, health and emotional state, racial or ethnic origin, geographical background, social identity, and socio-economic status~\cite{Nautsch-PreservingPrivacySpeech-CSL-2019}. Speaker recognition systems can also reveal the speaker's identity.
Formed in 2020, the VoicePrivacy initiative~\cite{tomashenko2020introducing} is spearheading the effort to develop privacy preservation solutions for speech technology. We aim to foster progress in the development of anonymization and pseudonymization solutions which suppress personally identifiable information contained within recordings of speech while preserving linguistic content, paralinguistic attributes, intelligibility and naturalness.
VoicePrivacy takes the form of a competitive benchmarking challenge, with common datasets, protocols and metrics. The first edition of VoicePrivacy was held in 2020 \cite{tomashenko2020introducing,Tomashenko2021CSl,Tomashenko2021CSlsupplementay,bonastre2021benchmarking}.
VoicePrivacy 2022, the second edition, starts in March 2022 and culminates in the VoicePrivacy Challenge workshop held in conjunction with the 2nd  Symposium on Security and Privacy in Speech Communication (SPSC)\footnote{\label{fn:spsc}2nd Symposium on Security and Privacy in Speech Communication: https://symposium2022.spsc-sig.org/}, a joint 
event co-located with INTERSPEECH 2022\footnote{https://www.interspeech2022.org/} in Incheon, Korea.

\emph{Anonymization} refers to the goal of suppressing personally identifiable information in the speech signal, leaving other attributes intact \cite{Tomashenko2021CSl}.
Note that, in the legal community, the term ``anonymization'' means that this goal has been achieved. Here, it refers to the task to be addressed, even when the method being evaluated has failed.
Anonymization requires altering not only the speaker's voice, but also linguistic content, extralinguistic traits, and background sounds which 
might reveal the speaker's identity.
As a step towards this goal, 
and in keeping with the inaugural VoicePrivacy 2020 Challenge, the second edition focuses on \textit{voice anonymization}, that is the task of altering the speaker's voice to conceal their identity to the greatest possible extent, while leaving the linguistic content and paralinguistic attributes intact.

 This document describes plans for the challenge, the datasets, protocol and the set of metrics that will be used for assessment, in addition to evaluation rules and guidelines for registration and submission.

\section{Challenge objectives}
The grand objective of the VoicePrivacy Challenge is to foster progress in the development of anonymization techniques for speech data. The specific technical goals are summarised as follows. They are to:

\begin{itemize}
\item facilitate the development of novel techniques which suppress speaker-discriminative information within speech signals;
\item promote techniques which provide effective anonymization while protecting linguistic content, paralinguistic attributes, intelligibility and naturalness;
\item provide a level playing field to facilitate the comparison of different anonymization solutions using a common dataset and protocol;
\item investigate metrics for the evaluation and meaningful comparison of different anonymization solutions.
\end{itemize}

VoicePrivacy participants will be provided with common datasets, protocols and a suite of software packages that will enable them to evaluate anonymization performance.

\section{Task}
\label{sec:task}

Privacy preservation is formulated as a game between \emph{users} who 
share some data and \emph{attackers} who access this data or data derived from it and wish to infer information about the users \cite{qian2018towards,srivastava2019evaluating,tomashenko2020introducing}. To protect their privacy, the users share data that contain as little personal information as possible while still allowing one or more downstream goals to be achieved. To infer personal information, the attackers may use additional prior knowledge.

Focusing on speech data, a given privacy preservation scenario is specified by:
(i)~the nature of the data: waveform, features, etc.; (ii)~the information seen as personal: speaker identity, traits, linguistic content, etc.; (iii)~the downstream goal(s): human communication, automated processing, model training, etc.; (iv)~the data accessed by the attackers: one or more utterances, derived data or models, etc.; (v)~the attackers' prior knowledge: previously collected speech data, the applied privacy preservation method, etc.
Different specifications lead to different privacy preservation methods from the users' point of view and different attacks from the attackers' point of view.

Here, we consider the scenario where speakers want to hide their identity to the greatest possible extent while allowing the desired downstream goals to be achieved, while attackers want to identify the speakers from their utterances.

\subsection{Anonymization task}
\label{subsec:user_goals}

The utterances shared by the users are referred to as \emph{trial} utterances. In order to hide his/her identity, each user passes these utterances through a voice anonymization system prior to sharing. The resulting utterances sound as if they were uttered by another speaker, which we refer to as a \emph{pseudo-speaker}. The pseudo-speaker might, for instance, be an artificial voice not corresponding to any real speaker.

The task of challenge participants is to develop this anonymization system. 
It should: 
\begin{enumerate}[label=(\alph*)]
    \item output a speech waveform; 
    \item conceal the speaker identity; 
    \item leave the linguistic content and paralinguistic attributes unchanged; 
    \item ensure that all trial utterances from a given speaker are uttered by the same pseudo-speaker, while trial utterances from different speakers are uttered by different pseudo-speakers.\footnote{We refer to this type of anonymization as \textit{speaker-level} anonymization.
An alternative approach is \textit{utterance-level} anonymization where different utterances of the same source speaker are anonymized using different parameters of the anonymization system, so that they may sound as if they were uttered by different pseudo-speakers. For evaluation, we assume that only \textit{speaker-level} anonymization should be applied to trial and enrollment data, while \textit{utterance-level} anonymization will be applied to the training data for training strong attack models.}
\end{enumerate}

The requirement (c) promotes the achievement of all possible downstream goals to the greatest possible extent. In practice, we restrict ourselves 
two use cases: automatic speech recognition (ASR)  training and/or decoding, and multi-party human conversations.
The achievement of these goals is assessed via a range of \emph{utility} metrics.
Specifically, we will measure ASR performance using a model trained on 
anonymized data.
In addition, the pitch correlation $\boldsymbol{\rho}^{F_0}$ between original (unprocessed) and anonymized speech signals 
will be used 
as a secondary objective utility metric to measure the degree to which intonation is preserved in anonymized speech, and subjective speech intelligibility and naturalness will also be measured.

The  requirement (d)  is motivated by the fact that, in a multi-party human conversation, the anonymized voices of all speakers must be distinguishable from each other and should not change over time. It will be assessed via the gain of voice distinctiveness $G_{\text{VD}}$ metric.

\subsection{Attack model}
\label{subsec:attack_model}


For each speaker of interest, the attacker is assumed to have access to one or more utterances spoken by that speaker. These utterances may or may not have been anonymized and are referred to as \textit{enrollment} utterances.

In this work, the
attackers have access to:
\begin{enumerate}[label=(\alph*)]
\item one or more anonymized trial utterances;
\item possibly, original or anonymized enrollment utterances for each speaker;
\item anonymized training data (and can retrain an automatic speaker verification system using this data).
\end{enumerate}

The protection of identity information is assessed via \emph{privacy} metrics, including objective and subjective speaker verifiability. These metrics assume different attack models.
The objective speaker verifiability metrics assume that the attacker has access to a single anonymized trial utterance and several anonymized enrollment utterances but, 
for subjective speaker verifiability evaluation, to a single anonymized trial utterance and a single original enrollment utterance.

\section{Data}\label{sec:data}
Several publicly available corpora are used for the training, development and evaluation of voice anonymization systems.  They are the same as for the VoicePrivacy 2020 Challenge~\cite{tomashenkovoiceprivacy} and comprise subsets from the following corpora:
\begin{itemize}
    \item \textit{\textbf{LibriSpeech}}\footnote{\label{fn:url1}Librispeech: \url{http://www.openslr.org/12}}~\cite{panayotov2015librispeech} is a corpus of read English speech derived from audiobooks and designed for
ASR research. It contains approximately \numprint{1000}~hours of speech sampled at 16~kHz.

    \item \textit{\textbf{LibriTTS}}\footnote{LibriTTS: \url{http://www.openslr.org/60/}}~\cite{zen2019libritts} is a corpus of English speech derived from LibriSpeech and designed for research in text-to-speech (TTS). It contains approximately 585~hours of read English speech sampled at 24~kHz.

    \item {\textbf{\textit{VCTK}}}\footnote{VCTK, release version 0.92: \url{https://datashare.is.ed.ac.uk/handle/10283/3443}}~\cite{yamagishi2019cstr} is a corpus of read speech collected from 109 native speakers of English with various accents. It was originally aimed for research in TTS and contains approximately 44~hours of speech sampled at 48~kHz.
    
    \item \textit{\textbf{VoxCeleb-1,2}}\footnote{VoxCeleb: \url{http://www.openslr.org/60/}}~\cite{nagrani2017voxceleb,chung2018voxceleb2} is an audiovisual corpus extracted from videos uploaded to YouTube and designed for speaker verification research. It contains
    approximately \numprint{2770}~hours of speech sampled at 16~kHz collected from \numprint{7363}~speakers, covering a wide range of accents and languages.

\end{itemize}

A detailed description of the datasets provided for training, development and evaluation is presented below and in Table~\ref{tab:data}.

\paragraph{Training set --}
The training set comprises the 
\textit{VoxCeleb-1,2} corpus \cite{nagrani2017voxceleb,chung2018voxceleb2} and 
subsets of the \textit{{LibriSpeech}} \cite{panayotov2015librispeech} and \textit{LibriTTS} \cite{zen2019libritts} corpora. 
The selected subsets are detailed in Table~\ref{tab:data} (top).

\begin{table}[!t]
\centering
  \caption{Number of speakers and utterances in the training, development, and evaluation sets~\cite{tomashenko2020introducing}.}\label{tab:data}
 \resizebox{0.95\textwidth}{!}{
  \centering
  \begin{tabular}{|c|l|l|r|r|r|r|}
\hline
 \multicolumn{3}{|l|}{\textbf{Subset}} &  \textbf{Female} & \textbf{Male} & \textbf{Total} & \textbf{\#Utterances}  \\ \hline \hline
\multirow{5}{*}{\rotatebox{90}{Training~}} & \multicolumn{2}{l|}{VoxCeleb-1,2} & \numprint{2912} & \numprint{4451} & \numprint{7363} & \numprint{1281762} \\ \cline{2-7}
& \multicolumn{2}{l|}{LibriSpeech train-clean-100} & 125 & 126 & 251	& \numprint{28539} \\ \cline{2-7}
& \multicolumn{2}{l|}{LibriSpeech train-other-500} & 564 & 602 & \numprint{1166} & \numprint{148688}	\\ \cline{2-7}
& \multicolumn{2}{l|}{LibriTTS train-clean-100} & 123 & 124 & 247 & \numprint{33236} \\ \cline{2-7}
& \multicolumn{2}{l|}{LibriTTS train-other-500} & 560 & 600 & \numprint{1160} & \numprint{205044} \\ \hline\hline
\multirow{5}{*}{\rotatebox{90}{~ Development }} & LibriSpeech & Enrollment & 15 & 14 & 29 & 343\\ \cline{3-7}
& dev-clean & Trial & 20 & 20 & 40 & \numprint{1978}\\ \cline{2-7}
& & Enrollment & & & & 600 \\  \cline{3-3}\cline{7-7}
& VCTK-dev & Trial (different) & 15 & 15 & 30 & \numprint{10677} \\  \cline{3-3}\cline{7-7}
& & Trial (common) & & & &  695\\  \hline\hline
\multirow{5}{*}{\rotatebox{90}{Evaluation~}} & LibriSpeech & Enrollment & 16 & 13 & 29 & 438\\ \cline{3-7}
& test-clean & Trial & 20 & 20 & 40 & \numprint{1496}\\ \cline{2-7}
& & Enrollment & & & & 600 \\  \cline{3-3}\cline{7-7}
& VCTK-test & Trial (different) & 15 & 15 & 30 & \numprint{10748} \\ \cline{3-3}\cline{7-7}
& & Trial (common) & & & & 700 \\ \hline
\end{tabular}}
\end{table}
\normalsize

\paragraph{Development set --}
The development set comprises \textit{LibriSpeech dev-clean} and a subset of the VCTK corpus \cite{yamagishi2019cstr}, denoted  \textit{VCTK-dev} (see Table~\ref{tab:data}, middle). Both are split into trial and enrollment subsets. For \textit{LibriSpeech dev-clean}, speakers in the enrollment set are a subset of those in the trial set. 
For \textit{VCTK-dev}, we use the same speakers for enrollment and trial and consider two trial subsets: \textit{common} and \textit{different}. The \textit{common} subset comprises utterances $\#1-24$ in the VCTK corpus that are identical for all speakers. This choice is intended to support subjective evaluation of speaker verifiability in a text-dependent manner. The enrollment and \textit{different} subsets comprise distinct utterances for all speakers.

\paragraph{Evaluation set --} Similarly, the evaluation set comprises \textit{LibriSpeech test-clean} and a subset of VCTK, denoted \textit{VCTK-test} (see Table~\ref{tab:data}, bottom).

\section{Utility and privacy metrics}
\label{sec:perf}

We consider objective and subjective privacy metrics to assess speaker verifiability. We also propose objective and subjective utility metrics to assess fulfillment of the user goals specified in Section~\ref{sec:task}.

\subsection{Objective assessment of the privacy-utility tradeoff}
\label{sec:perf_objective}
A pair of metrics will be used for the objective ranking of submitted systems: the equal error rate (EER) as the privacy metric and the word error rate (WER) as the primary utility metric. 
These metrics rely on automatic speaker verification (ASV) and automatic speech recognition (ASR) systems, both trained on the \textit{LibriSpeech-train-clean-360} dataset, statistics for which are presented in Table~\ref{tab:train-eval-metrics}. 
Training and evaluation will be performed with the provided recipes.\footnote{Evaluation scripts: \url{https://github.com/Voice-Privacy-Challenge/Voice-Privacy-Challenge-2022}}

\begin{table}[htbp]
  \caption{Statistics of the training dataset for the objective evaluation systems.}\label{tab:train-eval-metrics}
  \renewcommand{\arraystretch}{1.1}
  \centering
  \begin{tabular}{|l|r|r|r|r|r|}
\Xhline{0.6pt}
   \multirow{2}{*}{\textbf{Subset}} &   \multirow{2}{*}{\textbf{Size,h}} & \multicolumn{3}{c|}{\textbf{Number of Speakers}} &  \multirow{2}{*}{\textbf{Number of Utterances}} \\ \cline{3-5}
  &  & \textbf{Female} & \textbf{Male} & \textbf{Total} & \\ \hline \hline
  LibriSpeech: train-clean-360 & 363.6 & 439 & 482	 &	921	& \numprint{104014}	\\ \Xhline{0.6pt}
  \end{tabular}
\end{table}

New to the 2022 edition is the use of multiple evaluation conditions specified with a set of minimum target privacy requirements.  Submissions to each condition that meet each minimum target privacy requirement will then be ranked according to their protection of utility.  The goal is to measure the privacy-utility trade-off of any given solution at multiple operating points, e.g.\ when they are configured to offer better privacy at the cost of utility and vice versa.  This approach to assessment aligns better the VoicePrivacy Challenge with the user expectation of privacy and allows for a more comprehensive evaluation of each solution, while also providing participants with a set of clear optimisation criteria.  The privacy and primary utility metrics will be used for this purpose.

\begin{figure}[bth!]
\centering\includegraphics[width=115mm]{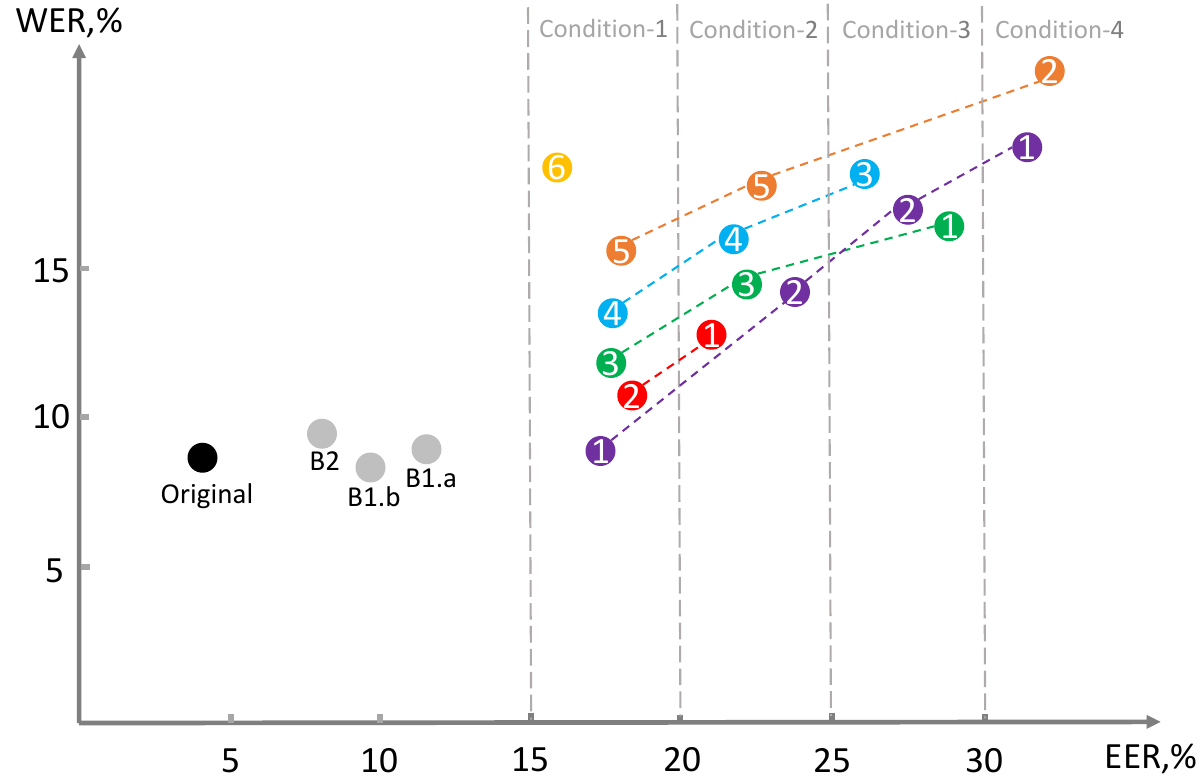}
\caption{Example system rankings according to the privacy (EER) and utility (WER) results for 4  EER threshold conditions. Different colors correspond to 6 different teams.  Numbers within each circle show system ranks for a given condition. Grey circles correspond to the baseline systems, and a black one -- to the original (unprotected) system.}
\label{fig:thresholds}
\end{figure}

Minimum target privacy requirements  
are specified with a set of $N$ minimum target EERs: \{EER$_1$, \ldots, EER$_N$\}. 
Each minimum target EER constitutes a separate evaluation condition.
Participants are encouraged to submit solutions to as many conditions as possible.  Submissions to any one condition $i$ should achieve an average EER on the VoicePrivacy 2022 test datasets greater than the corresponding minimum EER$_i$.
The average EER is computed by averaging the three EERs obtained on \textit{LibriSpeech-test-clean}, \textit{VCTK-test (common)} and \textit{VCTK-test (different)} with weights of 0.5, 0.1 and 0.4, respectively.\footnote{These weights assign the same importance to LibriSpeech and VCTK, and account for the different number of trials in the two VCTK subsets.}
The set of valid submissions for each minimum EER$_i$ will then be ranked according to the corresponding average WER results computed by averaging the two WERs obtained on \textit{LibriSpeech-test-clean} and \textit{VCTK-test} with equal weights. The VoicePrivacy 2022 Challenge involves $N=4$ conditions with minimum target EERs of: EER$_1=15\%$, EER$_2=20\%$,
EER$_3=25\%$, EER$_4=30\%$.
The lower the WER 
for a given EER condition, the better the rank of the considered system.
A depiction of example results and system rankings according to this methodology is shown in Figure~\ref{fig:thresholds}.

\subsubsection{Privacy metric: equal error rate (EER)}\label{sec:asv-eval}

\begin{figure}[b!]
\centering\includegraphics[width=153mm]{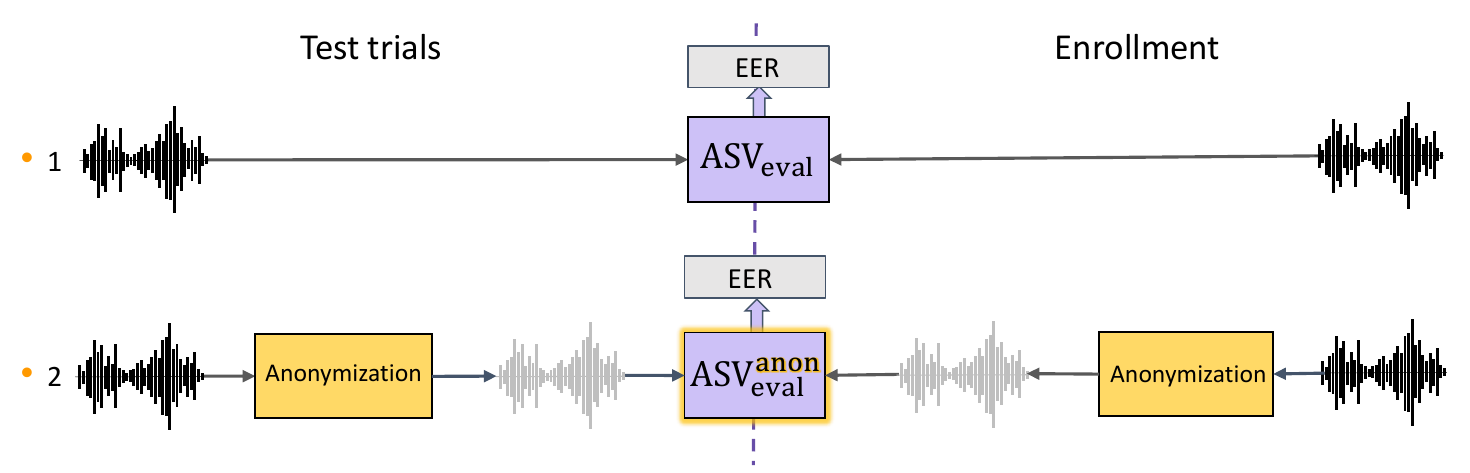}
\caption{ASV evaluation (1) \textit{unprotected}: original trial and enrollment data, $ASV_\text{eval}$ trained on original data;
(2)~\textit{semi-informed} attacker: \textit{speaker-level} anonymized trial and enrollment data with different pseudo-speakers,  $ASV_\text{eval}^{\text{anon}}$ trained on \textit{utterance-level} anonymized data.}
\label{fig:asv-eval}
\end{figure}

The evaluation of objective speaker verifiability assumes that the attacker has access to one trial utterance and several enrollment utterances.
The ASV system is based on x-vector speaker embeddings and probabilistic linear discriminant analysis (PLDA) \cite{snyder2018x}. For every pair of enrollment and trial x-vectors, it outputs a log-likelihood ratio (LLR) score from which a same-speaker vs.\ different-speaker decision is made by thresholding.
Denoting by $P_\text{fa}(\theta)$ and $P_\text{miss}(\theta)$ the false alarm and miss rates at threshold~$\theta$, the EER metric corresponds to the threshold $\theta_\text{EER}$ at which the two detection error rates are equal, i.e., $\text{EER}=P_\text{fa}(\theta_\text{EER})=P_\text{miss}(\theta_\text{EER})$.

As seen in Figure~\ref{fig:asv-eval}, this metric is computed for two evaluation scenarios~\cite{Tomashenko2021CSl,srivastava2021}:
\begin{enumerate}\setlength\itemsep{0.05em}
\item \textit{Unprotected} --- No anonymization is performed by users. The attacker has access to original (i.e., unprocessed) trial and enrollment data and uses an ASV system (denoted $ASV_\text{eval}$) trained on the original \textit{LibriSpeech-train-clean-360} data.
\item \textit{Semi-informed} \cite{maouche2021enhancing} --- 
Users anonymize their trial data. The attacker has access to original enrollment data and to the anonymization system, which is assumed to be publicly available. Using that system, the attacker performs \textit{speaker-level} anonymization of the enrollment data so as to reduce the mismatch.
The trial and enrollment data are anonymized using different pseudo-speakers, since the attacker does not know the pseudo-speaker chosen by each user.
In addition, the attacker uses that system for \textit{utterance-level} anonymization of the \textit{LibriSpeech-train-clean-360} dataset and retrains the ASV system (now denoted $ASV_\text{eval}^{\text{anon}}$) on it. We found the latter to lead to a lower EER, i.e., a stronger attack, than \textit{speaker-level} anonymization of the ASV training set \cite[Table~V]{shamsabadi2022dp}.
This attack model is actually the strongest known to date, hence we consider it as the most reliable for privacy assessment.\footnote{In the VoicePrivacy 2020 Challenge, evaluation relied on three attack models referred to as \textit{ignorant}, \textit{lazy-informed}, and \textit{semi-informed}, corresponding to attackers with different knowledge \cite{Tomashenko2021CSl,srivastava2021,tomashenko2020posteval}. The \textit{semi-informed} attack model differed from the one considered here, since it assumed \textit{speaker-level} anonymization of the ASV training dataset.}
\end{enumerate}

The number of same-speaker and different-speaker trials in the development and evaluation datasets is given in Table~\ref{tab:trials}. For a given speaker, all enrollment utterances 
are used to compute an average x-vector for enrollment.

\begin{table}[htbp]
  \caption{Number of speaker verification trials.}\label{tab:trials}
  \renewcommand{\tabcolsep}{0.13cm}
  \centering
   \resizebox{0.87\textwidth}{!}{
  \begin{tabular}{|l|l|l|r|r|r|}
\hline
 \multicolumn{2}{|l|}{\textbf{Subset}} & \textbf{Trials} &  \textbf{Female} & \textbf{Male} & \textbf{Total}  \\ \hline \hline
\multirow{6}{*}{\rotatebox{90}{Development~}} & LibriSpeech & Same-speaker & 704 & 644 & \numprint{1348} \\ \cline{3-6}
 & dev-clean & Different-speaker	& \numprint{14566} & \numprint{12796} &	\numprint{27362} \\ \cline{2-6}
& \multirow{4}{*}{VCTK-dev} & Same-speaker (different) & \numprint{1781}	& \numprint{2015} & \numprint{3796} \\ \cline{3-6}
 & & Same-speaker (common) & \numprint{344} & \numprint{351} & \numprint{695} \\  \cline{3-6}
 & & Different-speaker (different) & \numprint{13219} & \numprint{12985} & \numprint{26204} \\ \cline{3-6}
 & & Different-speaker (common) & \numprint{4810} &	\numprint{4911} & \numprint{9721} \\ \hline\hline
\multirow{6}{*}{\rotatebox{90}{Evaluation~}} & LibriSpeech & Same-speaker & 548 & 449	& \numprint{997} \\ \cline{3-6}
  & test-clean & Different-speaker & \numprint{11196} & \numprint{9457} &	\numprint{20653} \\ \cline{2-6}
 & \multirow{4}{*}{VCTK-test} & Same-speaker  (different) & \numprint{1944} & \numprint{1742} & \numprint{3686} \\ \cline{3-6}
 & & Same-speaker (common) & \numprint{346} & \numprint{354} & \numprint{700} \\  \cline{3-6}
 & & Different-speaker (different) & \numprint{13056} &	\numprint{13258} &\numprint{26314} \\ \cline{3-6}
 & & Different-speaker (common) & \numprint{4838} & \numprint{4952} & \numprint{9790}  \\ \hline
  \end{tabular}}
\end{table}
\normalsize

\subsubsection{Primary utility metric: word error rate (WER)}\label{sec:wer}

The ability of the anonymization system to preserve linguistic information is assessed using an ASR system based on the Kaldi toolkit \cite{povey2011kaldi}. We adapted the Kaldi recipe for LibriSpeech involving an acoustic model with  a factorized time delay neural network (TDNN-F) architecture \cite{povey2018semi,peddinti2015time} and a \textit{large} trigram language model.
The ASR system outputs a word sequence and the WER is calculated as
\begin{equation*}
\text{WER}=\frac{N_\text{sub}+N_\text{del}+N_\text{ins}}{N_\text{ref}},
\end{equation*}
where $N_\text{sub}$, $N_\text{del}$, and $N_\text{ins}$, are the number of substitution, deletion, and insertion errors respectively, and $N_\text{ref}$ is the number of words in the reference. %
The lower the WER, the greater the utility.

\begin{figure}[h!]
\centering\includegraphics[width=95mm]{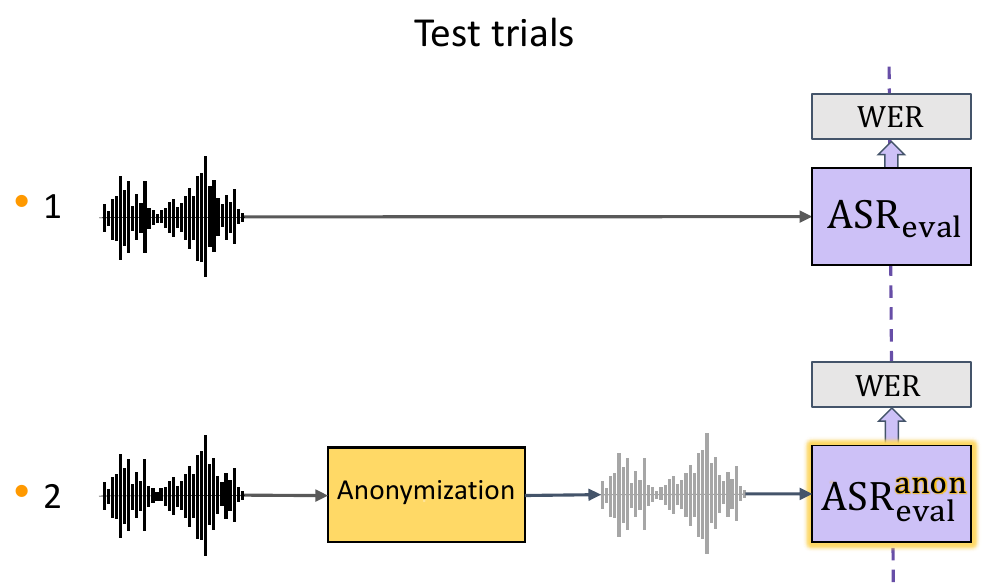}
\caption{ASR evaluation (1) original data decoded with \textrm{$ASR_\text{eval}$} trained on original data; 
(2)~\textit{speaker-level} anonymized data decoded with \textrm{$ASR_\text{eval}^\text{anon}$} trained on  \textit{utterance-level} anonymized  data.
WER is computed on the \textit{trial} utterances of the development and evaluation datasets.}
\label{fig:asr-eval}
\end{figure}

As shown in Figure~\ref{fig:asr-eval}, we consider two ASR evaluation scenarios: (1) the original
trial data is decoded using the ASR model (denoted \textrm{$ASR_\text{eval}$}) trained on the original \textit{LibriSpeech-train-clean-360} dataset, while
(2) anonymized trial data is decoded using the ASR model (denoted \textrm{$ASR_\text{eval}^\text{anon}$}) trained on the anonymized (\textit{utterance-level}) \textit{LibriSpeech-train-clean-360} dataset. As demonstrated in \cite{tomashenko2020posteval,Tomashenko2021CSl}, the latter significantly decreases the WER on anonymized data compared to decoding with the \textrm{$ASR_\text{eval}$} model trained on original data.

\subsection{Secondary utility metrics}
In addition to the primary metrics, we consider two secondary utility metrics, namely pitch correlation $\rho^{F_0}$ and
the gain of voice distinctiveness $G_{\text{VD}}$.
While these secondary metrics are not used for ranking, all submissions are expected to exceed a minimum pitch correlation threshold.

\subsubsection{Pitch correlation metric $\boldsymbol{\rho}^{F_0}$}\label{subsubsec:f0}

The new pitch correlation metric has been introduced to
provide a measure of how well anonymization preserves the intonation of the original utterance.  
Intonation is among \emph{other speech attributes} that should remain intact following anonymisation. 

The pitch correlation metric $\boldsymbol{\rho}^{F_0}$ is 
the Pearson correlation between the pitch sequences, estimated according to~\cite{Hirst07apraat}, of original and anonymized utterances. 
It is computed as follows.  
Pitch sequences are estimated for each pair of utterances and the shortest sequence is linearly interpolated so that its length matches that of the longest sequence.  
The temporal lag between original and anonymized utterances is then adjusted in order to maximise the Pearson cross-correlation.
The latter is estimated using only segments where both original and anonymized utterances are voiced. 
Estimates of $\boldsymbol{\rho}^{F_0}$, calculated for development and evaluation datasets, are averages of the pitch correlation values for all utterances in each dataset.

While a secondary metric, \textbf{all} submissions should achieve a minimum average pitch correlation of $\rho^{F_0}>0.3$ \textbf{for each dataset and for each condition}.  Solutions that achieve lower average pitch correlations will be considered invalid.  The threshold was set according to results derived from arbitrary anonymisation solutions that do not preserve intonation, e.g.\ ASR+TTS solutions.  The threshold is a modest minimum correlation; all baseline solutions achieve average pitch correlations in the order of 0.7 hence submissions that make a reasonable attempt to preserve intonation should achieve correlation well above the minimum threshold.

\subsubsection{Gain of voice distinctiveness $\text G_\text{VD}$}\label{subsec:gainvd}

The gain of voice distinctiveness metric aims to evaluate the requirement to preserve voice distinctiveness.
It relies on voice similarity matrices~\cite{noe2020speech,noe2021csl}.

A voice similarity matrix $M=( M(i,j))_{1 \le i \le N,1 \le j \le N}$ is defined for a set of $N$ speakers using similarity values $ M(i,j)$ computed for speakers $i$ and $j$ as follows:
\begin{equation}
\small
     M(i,j) = \mathrm{sigmoid}\left({\frac{1}{n_{i}n_{j} }
    \displaystyle\sum_{\substack{1 \le k \le n_{i} \text{ and } 1 \le l \le n_{j} \\ k\neq l \text{ if } i=j } }{\text{LLR}(x^{(i)}_{k},x^{(j)}_{l})}}\right)
    \label{equ:: M}
\end{equation}
where $\text{LLR}(x^{(i)}_{k},x^{(j)}_{l})$ is the log-likelihood-ratio obtained by comparing the $k$-th segment from the $i$-th speaker with the $l$-th segment from the $j$-th speaker, and where $n_{i}$ and $n_j$ are the numbers of segments for each speaker. 
Two matrices are computed using the $ASV_\text{eval}$ model trained on original data:
$M_\text{oo}$ on original data and $M_\text{aa}$ on anonymized data.
For each of these two similarity matrices, the diagonal dominance $D_\text{diag}(M)$ is computed as the absolute difference between the mean values of diagonal and  off-diagonal elements:
\begin{equation}
\small
    D_{\text{diag}}(M)\hspace{-0.7mm}=\hspace{-0.7mm}\displaystyle
    \Bigg|
    \sum_{1\leq i \leq N} \frac{ M(i,i)}{N}
    \displaystyle
    - 
    \sum_{\substack{1 \le j \le N \text{ and } 1 \le k \le N \\j \neq k}}
    \frac{ M(j,k)}{N(N-1)}
    \Bigg|.
    \label{eq:ddiag}
\end{equation}
The gain of voice distinctiveness metric ($G_{\text{VD}}$) \cite{noe2020speech} is defined as the ratio of diagonal dominance of the two matrices:
\begin{equation}
 G_{\text{VD}} = 10\log_{10}  \frac{D_\text{diag}(M_\text{aa})}{D_\text{diag}(M_\text{oo})},   
\end{equation}
where a gain of $G_{\text{VD}}=0$ implies that the voice distinctiveness remains the same on average after anonymization, and a gain above or below 0 corresponds respectively to an average increase or decrease in voice distinctiveness.

\subsection{Subjective metrics}\label{sec:subj-metr}
\label{sec:perf_subjective}

\begin{figure}[!b]
\centering
\includegraphics[width=0.95\linewidth]{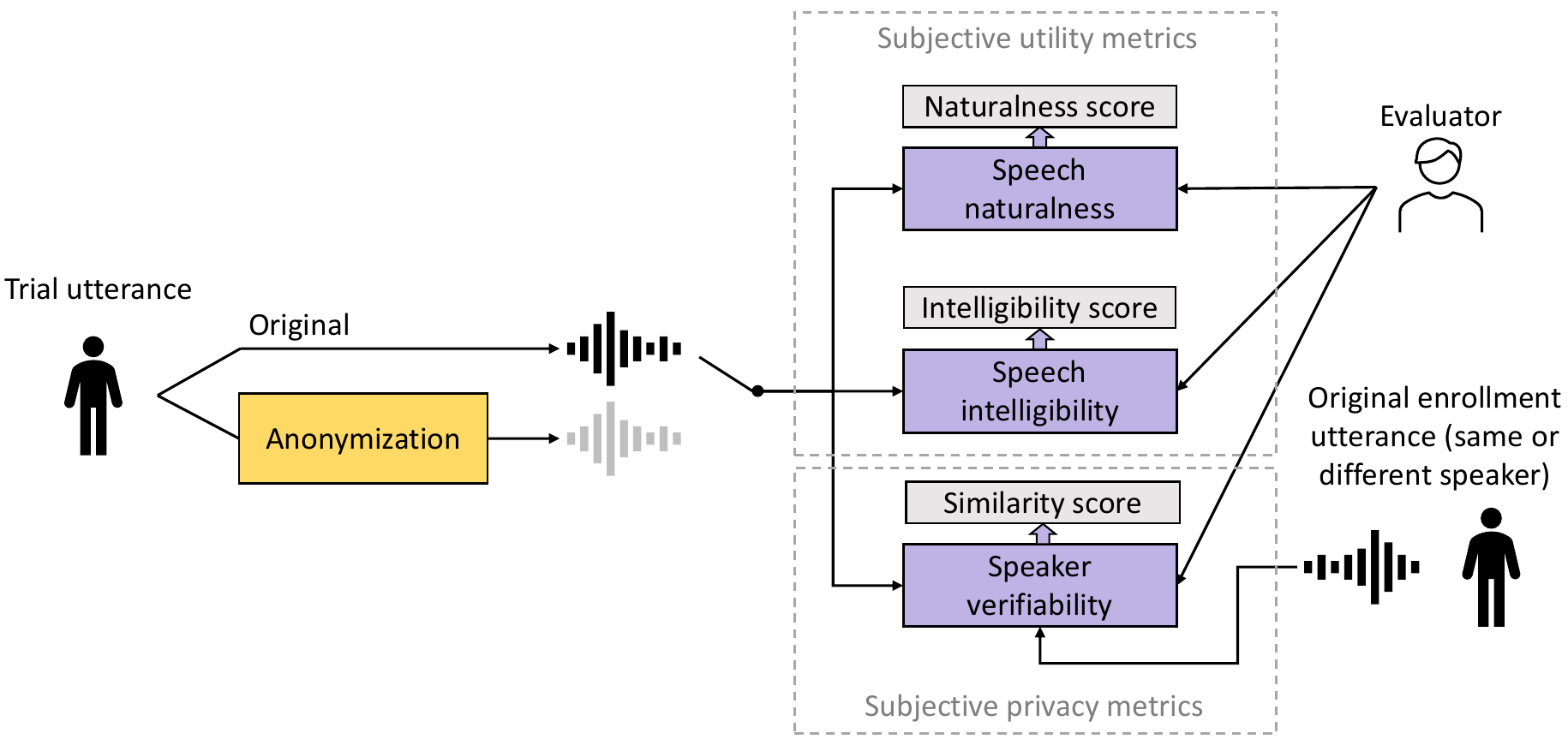}
\caption{Subjective evaluation test for speech naturalness, intelligibility, and speaker verifiability~\cite{Tomashenko2021CSl}.}
\label{fig:sub-eval-design}
\end{figure}

Subjective metrics 
include: 
speaker verifiability; speech intelligibility; speech naturalness.
They will each be evaluated via 
unified subjective evaluation tests carried
out by the organizers as described below and as
illustrated in Figure~\ref{fig:sub-eval-design}.  The approach is similar to that used for the VoicePrivacy 2020 Challenge~\cite{Tomashenko2021CSl}. 
For naturalness and intelligibility assessments, 
evaluators will be asked to rate a single
original or anonymized trial utterance at a time.
For naturalness, 
the evaluator will assign a score from 1 (`totally unnatural') to 10 (`totally natural'). 
For intelligibility, the evaluator
will assign
a score from 1 (`totally unintelligible') to 10 (`totally intelligible'). 
Assessments of speaker verifiability will be performed with pairs of utterances, an original enrollment utterance and an original or anonymized trial utterance collected from the same or a different speaker.
The evaluators will rate the similarity between the voices in enrollment and trial utterances using a scale of 1 to 10, where 1 denotes `different speakers' and 10 denotes `the same speaker'.
Evaluators will be instructed to assign scores through a role-playing game. When an evaluator starts an evaluation session, the following instruction is displayed:

\vspace{-10pt}

\small
\quotebox{Please imagine that you are working at a TV or radio company. You wish to broadcast interviews of person X, but this person X does not want to disclose his/her identity. Therefore you need to modify speech signals in order to hide it. You have several automated tools to change speaker identity. Some of them hide the identity well, but severely degrade audio quality. Some of them hide the identity, but the resulting speech sounds very unnatural and may become less intelligible. In such cases, the privacy of person X is protected, but you will receive many complaints from the audience and listeners of TV/radio programs. You need to balance privacy of person X and satisfaction of TV/radio program audience and listeners. Your task is to evaluate such automated tools to change speaker identity and find out well-balanced tools.}
\normalsize

Separate, detailed instructions are provided for the listening test involving each of the three subjective metrics. The evaluator is asked to imagine the described scene when evaluating the corresponding metric.

\small
\quotebox{
\textbf{Subjective speech intelligibility}

For the final task, you are required to listen to audio A again and try to understand the audio content. Please judge how understandable audio A is. 

You need to select one score between 1 and 10, where a higher score denotes higher intelligibility. In particular, 1 means ``audio A is NOT understandable at all'' and 10 means ``audio A is perfectly understandable.
}
\normalsize
\small
\quotebox{
\textbf{Subjective speech naturalness}

You will listen to either original audio and audio modified by the above anonymization tools. Some of them result in artifacts and degradation due to poor audio processing. 

Now, please listen to audio A and answer how much you can hear the audio degradation. Please judge based on the characteristics of the audio rather than what is being said. 

You need to select a score between 1 and 10, where a higher score indicates less degradation. In particular, 1 means ``audio A exhibits severe audio degradation'' and 10 means ``audio A does not exhibit any degradation''. Please note that the original audio includes background noise.}
\normalsize
\small
\quotebox{
\textbf{Subjective speaker verifiability}

Your next task is to compare the processed or unprocessed audio A with audio B. From the voices, you must determine whether they are from the same person or another person.

Now, please listen to audio A above and audio B below, and determine if they were uttered by the same speaker. Please judge based on the characteristics of the voice rather than what is being said.

You need to select one score between 1 and 10, where a higher score denotes higher speaker similarity. In particular, 1 means ``audio A and B were uttered by different speakers for sure'' and 10 means ``audio A and B were uttered by the same speaker for sure.
}
\normalsize

By using clean speech of the same original speaker or a different speaker as Sample A, we will have anchors in the listening test and can visualize the performance of each participant system through detection error tradeoff (DET) curves~\cite{martin1997det} as described in \cite{Tomashenko2021CSl}.
 These curves assume a detection
task, where the decision for a given trial is made by comparing the score with
a threshold. The false alarm and miss rates are computed as a function of the
threshold and plotted against each other. For naturalness and intelligibility the
task is to detect original data, while for speaker similarity the task is to detect
whether the trial utterance is from the same speaker as the enrollment utterance. The closer the DET curves are to the top-right corner of each plot, the higher the naturalness, intelligibility, and privacy preservation.

\section{Baselines}\label{sec:baseline}

Three different baseline systems have been developed for the challenge.\footnote{\label{fn:scripts}All baseline systems are available online:  \url{https://github.com/Voice-Privacy-Challenge/Voice-Privacy-Challenge-2022}}

\subsection{Anonymization  using x-vectors and neural waveform models: B1.a and B1.b}

The baseline anonymization systems \textbf{B1.a} and \textbf{B1.b} are based on a common approach to x-vector modification and on two different speech synthesis components.

\subsubsection{B1.a}\label{subsec:b1a}
 
\begin{figure}[t]
\centering\includegraphics[width=110mm]{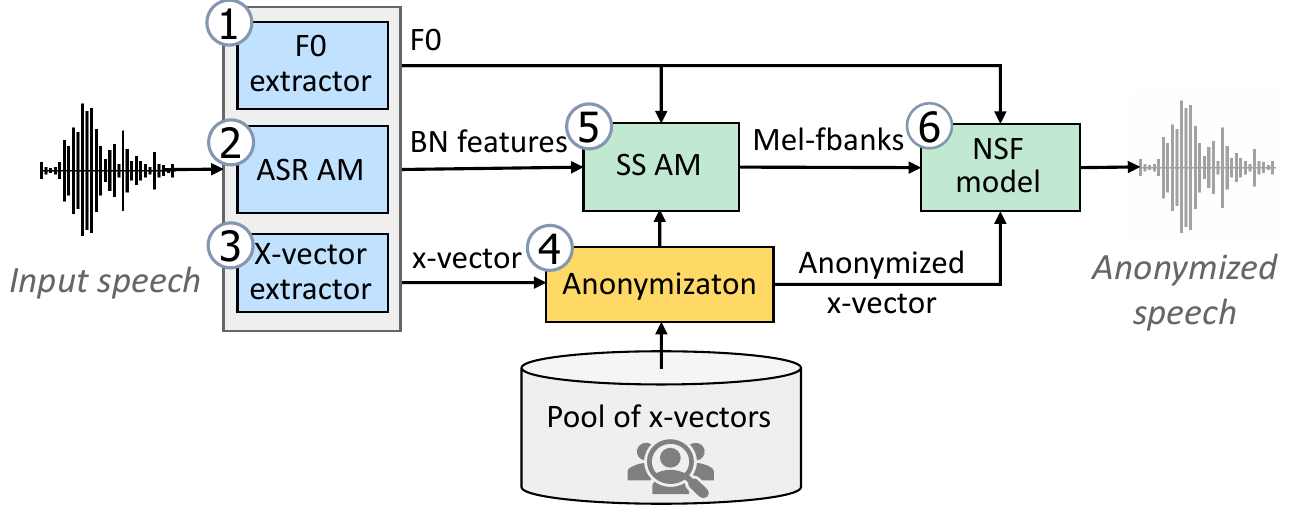}
\caption{First baseline anonymization system \textbf{B1.a}~\cite{tomashenko2020introducing}. 
}
\label{fig:baseline1.a}
\end{figure}

The first baseline \textbf{B1.a} is the primary baseline  of the VoicePrivacy 2020 Challenge~\cite{tomashenko2020introducing}. 
It is based on the voice anonymization method proposed in~\cite{fang2019speaker} and shown in Figure~\ref{fig:baseline1.a}.
Anonymization is performed in three steps:
\begin{itemize}
\item \textbf{Step 1 -- Feature extraction:} extraction of the speaker x-vector~\cite{snyder2018x}, the fundamental frequency (F0) and 
bottleneck (BN) features
from the original audio waveform.
\item \textbf{Step 2 -- X-vector anonymization:} anonymization of the source-speaker x-vector using an external pool of speakers.

\item \textbf{Step 3 -- Speech synthesis:} synthesis of a speech waveform from the
anonymized x-vector and the original BN and F0 features using an acoustic model and a neural waveform model.
\end{itemize}

\begin{table}[b!]
  \caption{Modules and training corpora for the  anonymization systems \textbf{B1.a} and \textbf{B1.b}. The module indexes are the same as in Figures~\ref{fig:baseline1.a} and \ref{fig:baseline1.b}. Superscript numbers  represent feature dimensions.}\label{tab:data-baseline-models}
 \renewcommand{\tabcolsep}{0.05cm} 
 \renewcommand{\arraystretch}{1.5}
  \centering
    \resizebox{0.98\textwidth}{!}{
  \begin{tabular}{|c|c|l|l|l|}
\Xhline{0.6pt}
  \textbf{\#} & \textbf{Module} & \textbf{Description} & \makecell[l]{\textbf{Output} \\ \textbf{features}} & \textbf{Data} \\ \hline \hline
  1 & \makecell{F0 \\extractor} &  pYAAPT\protect\footnotemark, uninterpolated  & F0$^{1}$ & - \\ \hline
  2 & \makecell[c]{ASR \\ AM} & \makecell[l]{ TDNN-F\\  Input:  MFCC$^{40}$ + i-vectors$^{100}$ \\ 17 TDNN-F hidden layers \\ Output: 6032 triphone  ids \\ LF-MMI and CE criteria \\ }  & \makecell[l]{ BN$^{256}$ \textrm{ features} \\   extracted from \\ the final hidden \\ layer} &  \makecell[l]{LibriSpeech:\\ train-clean-100 \\ train-other-500} \\ \hline
3 & \makecell{X-vector \\ extractor} & \makecell[l]{TDNN \\ Input: MFCC$^{30}$ \\  7  hidden layers + 1 stats pooling layer \\ Output: 7232 speaker ids \\ CE criterion\\ } & \makecell[l]{ speaker \\ x-vectors$^{512}$} & VoxCeleb-1,2 \\ \hline
  4 & \multicolumn{2}{l|}{  ~~X-vector anonymization module} & \makecell[l]{ 
pseudo-\\speaker \\ x-vectors$^{512}$} &  \makecell[l]{ 
(Pool of \\speakers) \\ LibriTTS:  \\train-other-500} \\ \hline
\makecell[c]{5.a \\ \textcolor{mountainmeadow}{\textbf{B1.a}}} & \makecell{Speech\\ synthesis\\ AM} & \makecell[l]{Autoregressive (AR) network \\ Input: F0$^{1}$+ \textrm{BN}$^{256}$+ \textrm{x-vectors}$^{512}$ \\ FF * 2 + BLSTM + AR + LSTM * 2 \\ + highway-postnet \\ MSE criterion }  & \makecell[l]{ Mel-filterbanks$^{80}$} & \makecell[l]{LibriTTS: \\ train-clean-100} \\ \hline
\makecell[c]{5.b \\ \textcolor{mountainmeadow}{\textbf{B1.b}}} & \makecell{Speech\\ synthesis\\ AM} & \makecell[l]{sinc-hn-NSF in \cite{wang2019neural} + HiFi-GAN discriminators \cite{kong2020hifi} \\ Input: F0$^{1}$+ \textrm{BN}$^{256}$ + \textrm{x-vectors}$^{512}$ \\ Training criterion defined in Hifi-GAN \cite{kong2020hifi} }  & \makecell[l]{ speech waveform} & \makecell[l]{LibriTTS: \\ train-clean-100} \\ \hline
6 & \makecell{NSF \\  model} & \makecell[l]{sinc-hn-NSF in \cite{wang2019neural} \\ Input:  F0$^{1}$ + \textrm{Mel-fbanks}$^{80}$ + \textrm{x-vectors}$^{512}$  \\ STFT criterion} & speech waveform & \makecell[l]{LibriTTS: \\ train-clean-100} \\ \Xhline{0.6pt}
  \end{tabular}}
\end{table}
\footnotetext{pYAAPT: \url{http://bjbschmitt.github.io/AMFM_decompy/pYAAPT.html}}

     In order to implement these steps, four different models are required, as shown in Figure~\ref{fig:baseline1.a}. Details for training these components are presented in Table~\ref{tab:data-baseline-models}.

In \textit{Step 1}, to extract BN features, an ASR acoustic model (AM)  is trained (\#1 in Table~\ref{tab:data-baseline-models}). We assume that the BN features represent the linguistic content of the speech signal. The ASR AM has a factorized time delay neural network (TDNN-F)  model architecture~\cite{povey2018semi,peddinti2015time} and is trained using the Kaldi toolkit~\cite{povey2011kaldi}.
To encode speaker information, an x-vector extractor with a TDNN model topology (\#2 in Table~\ref{tab:data-baseline-models}) is also trained using Kaldi.

In \textit{Step 2}, for a given source speaker, a new anonymized x-vector is computed by averaging a set of candidate x-vectors from the speaker pool. 
Probabilistic linear discriminant analysis (PLDA) is used as a distance measure between these vectors and the x-vector of the source speaker.  
The candidate x-vectors for averaging are chosen in two steps. First, for a given source x-vector, the $N$ farthest x-vector candidates in the speaker pool are selected. Second, a smaller subset of $N^*$ candidates are chosen randomly among those $N$ vectors.\footnote{In the baselines \textbf{B1.a} and \textbf{B1.b}, the following parameter values are used: $N=200$ and $N^*=100$.}
The x-vectors for the speaker pool are extracted from a disjoint dataset (\textit{LibriTTS-train-other-500}).

In \textit{Step 3}, 
two modules are used to generate the speech
waveform: a speech synthesis AM that generates Mel-filterbank features
given the F0, the anonymized x-vector, and the BN features, and a neural source-filter (NSF) waveform model~\cite{wang2019neural} 
that produces a speech waveform given the F0, the anonymized x-vector, and the generated Mel-filterbank outputs.
Both models (\#5.a and \#6 in Table~\ref{tab:data-baseline-models}) are trained on the same corpus (\textit{LibriTTS-train-clean-100}).

More details about the baseline recipe can be found in the \href{https://github.com/Voice-Privacy-Challenge/Voice-Privacy-Challenge-2022}{provided scripts}\textsuperscript{\ref{fn:scripts}} and in~\cite{srivastava2020baseline,srivastava2021}.

\subsubsection{B1.b}\label{subsec:b1a}
\begin{figure}[h!]
\centering\includegraphics[width=110mm]{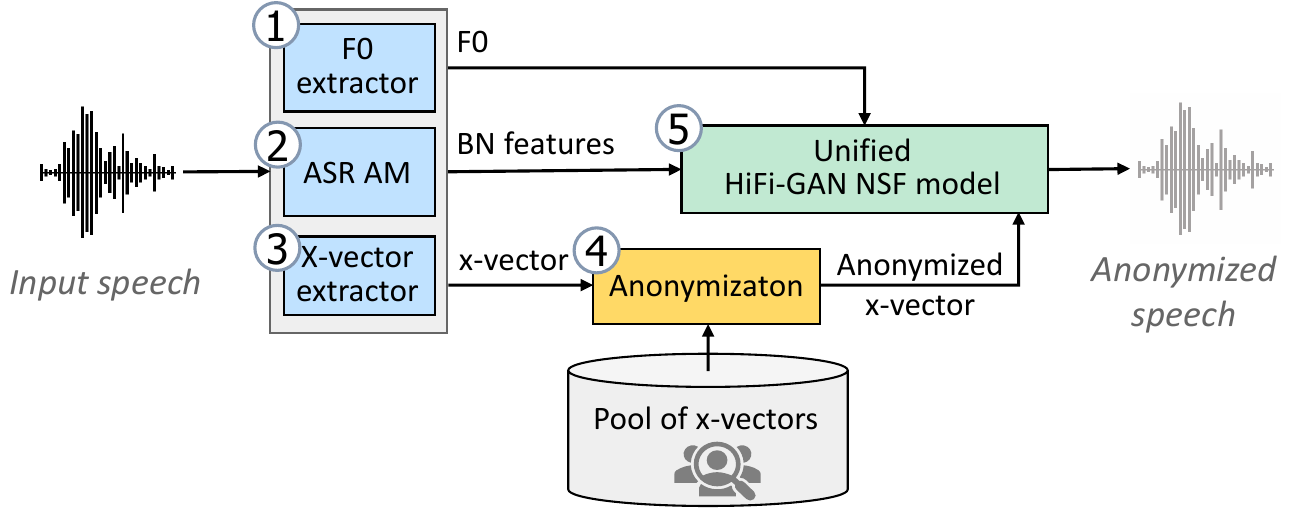}
\caption{Second baseline anonymization system \textbf{B1.b}. 
}
\label{fig:baseline1.b}
\end{figure}

The second baseline \textbf{B1.b}, shown in  Figure~\ref{fig:baseline1.b}, is based on the same idea as  \textbf{B1.a} and has the same x-vector extractor and anonymization modules, as well as pitch extractor and  ASR AM for linguistic feature extraction. The main difference between the two baselines is in the speech synthesis component (Step 3) of the anonymization system.
 While \textbf{B1.a} follows the traditional pipeline TTS approach and includes a speech synthesis AM and a separate waveform model, \textbf{B1.b} directly converts BN, F0, and x-vector features using an NSF model. It is unnecessary to generate Mel-filterbanks since BN-features already encode linguistic content. \textbf{B1.b} thus simplifies the system structure.

Another motivation for \textbf{B1.b} is to improve the quality of anonymized speech. Results from the VoicePrivacy 2020 Challenge \cite{Tomashenko2021CSl} indicate that speech anonymized by \textbf{B1.a} is inferior to \textbf{B2} in terms of subjective quality. One possible reason is the over-smoothing effect caused by the maximum-likelihood-based NSF training criterion ~\cite{wang2019neural}. One solution is to adopt a generative adversarial network (GAN)-based framework. As shown in Figure~\ref{fig:baseline1.b}, \textbf{B1.b} combines the NSF model (as the generator) with the discriminators of HiFi-GAN~\cite{kong2020hifi} and trains the model in the same manner as HiFi-GAN.  \textbf{B1.b} is trained using the same data as \textbf{B1.a}, namely \textit{LibriTTS-train-clean-100}. After training, the discriminators can be safely discarded, and only the trained NSF is used in the anonymization system.
The implementation of the TTS modules is based on Pytorch~\cite{Paszke2019Pytorch}.

\subsection{Anonymization using the McAdams coefficient: B2}

\begin{figure}[htp]
    \centering
    \includegraphics[width=115mm]{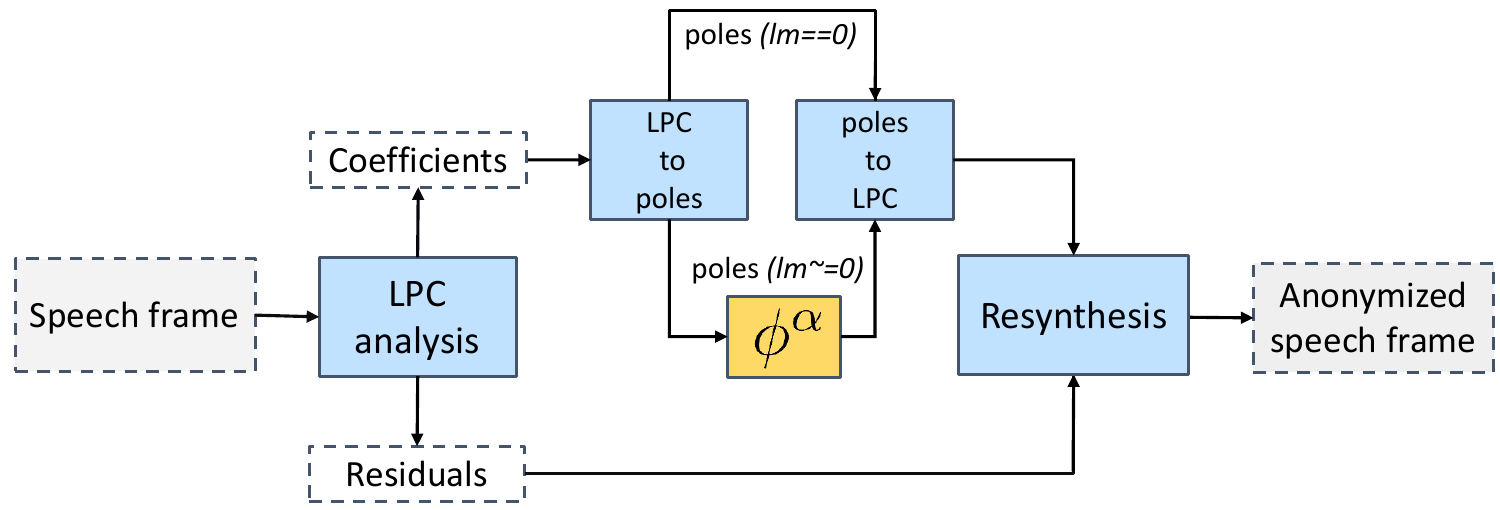}
    \caption{Third baseline anonymization system \textbf{B2}. 
    }
    \label{fig:lpc_processing}
\end{figure}

In contrast to \textbf{B1.a} and \textbf{B1.b}, the third baseline \textbf{B2} shown in Figure~\ref{fig:lpc_processing} does not require any training data and is based upon simple signal processing techniques. It employs the McAdams coefficient~\cite{mcadams1984spectral} to achieve anonymisation by shifting the pole positions derived from linear predictive coding (LPC) analysis of speech signals.

It starts with the application of frame-by-frame LPC source-filter analysis to derive LPC coefficients and residuals. The residuals are set aside for later resynthesis, whereas the LPC coefficients are converted into pole positions in the z-plane by polynomial root-finding. Each pole corresponds to a peak in the spectrum, resembling a formant position.
 The McAdams' transformation is applied to the phase of each pole: while real-valued poles are left unmodified, the phase $\phi$ (between 0 and $\pi$~radians) of poles with non-zero imaginary parts is raised to the power of the McAdams' coefficient $\alpha$ so that transformed poles have new, shifted phases of $\phi^\alpha$.
The value of $\alpha$ implies a contraction or expansion of the pole positions around $\phi=1$~radian.
For a sampling rate of 16~kHz, i.e.\ for the data used in this challenge, $\phi=1$~radian corresponds to approximately 2.5~kHz which is the approximate mean formant position~\cite{ghorshi2008cross}.
The corresponding complex conjugate poles are similarly shifted in the opposite direction  
and the new set of poles, including original real-valued poles, 
are converted back into LPC coefficients. Finally, the LPC coefficients and the residuals are used to resynthesise a new speech frame in the time domain.

The baseline \textbf{B2} is a randomized version of the anonymisation algorithm proposed in~\cite{patino2020speaker}, where the McAdams coefficient is sampled for each source speaker in the evaluation set from a uniform distribution: $\alpha\sim U(\alpha_\text{min},\alpha_\text{max})$.\footnote{This is different from the \textbf{B2} baseline of the 2020 challenge for which we used a constant value of $\alpha=0.8$.}
We use $\alpha_\text{min}=0.5$ and $\alpha_\text{max}=0.9$.

\subsection{Results}\label{subsec:results}

Results for the three baselines are reported in Tables~\ref{tab:asv-results} and \ref{tab:asr-results} in terms of the privacy (EER) and primary utility (WER)   metrics, and in Table~\ref{tab:second-results} in terms of the two secondary utility metrics.
EER and WER results for the three baseline systems are also depicted in Figure~\ref{fig:thresholds}.
\textbf{B1.b} has better WER than the other baseline systems, while \textbf{B1.a} demonstrates the highest average EER.

\begin{table*}[b!]
  \caption{Primary privacy evaluation: EER,\% achieved by $ASV_\text{eval}^\text{anon}$ on data processed by \textbf{B1.a}, \textbf{B1.b}, or \textbf{B2} vs.\ EER achieved by $ASV_\text{eval}$ on the original (Orig.) unprocessed data.}\label{tab:asv-results}
  \centering
  \begin{tabular}{|l|l|c||c||r|r|r|}
\Xhline{0.7pt}
 \multirow{2}{*}{	\textbf{Dataset}} &  \multirow{2}{*}{\textbf{Gender}} & \multirow{2}{*}{\textbf{Weight}} & \multicolumn{4}{c|}{\textbf{EER,\%}}\\ \cline{4-7}
& & &{	\textbf{Orig.}} & \textbf{B1.a} & \textbf{B1.b} & \textbf{B2~} \\ \hline \hline
\multirow{2}{*}{LibriSpeech-dev}	&	female	& 0.25 &	{8.67}	&	17.76	&	19.03	&	11.36	\\
	&	male 	& 0.25	&		1.24	&	6.37	&	5.59	&	1.40	\\ \hline
\multirow{2}{*}{VCTK-dev (different)}	&	female 	& 0.20	&		2.86	&	12.46	&	8.25	&	6.68	\\
	&	male 	& 0.20	&		1.44	&	9.33	&	6.01	&	6.35	\\  \hline
\multirow{2}{*}{VCTK-dev (common)}	&	female 	& 0.05	&		2.62	&	13.95	&	9.01	&	5.81	\\
	&	male 	& 0.05	&		{1.43}	&	13.11	&	9.40	&	8.83	\\  \hline
\multicolumn{3}{|c||}{\cellcolor{gray!7}Weighted average dev} &		\cellcolor{gray!7} 3.54	 & \cellcolor{gray!7}	11.74	& \cellcolor{gray!7} 9.93 	& \cellcolor{gray!7} 6.53 \\ \hline\hline
\multirow{2}{*}{LibriSpeech-test}	&	female 	& 0.25	&		7.66	&	12.04	&	9.49	&	7.12	\\
	&	male 	& 0.25	&		{1.11}	&	8.91	&	7.80	&	1.11	\\  \hline
\multirow{2}{*}{VCTK-test (different)}	&	female 	& 0.20	&		4.89	&	16.00	&	10.91	&	16.92	\\
	&	male 	& 0.20	&		{2.07}	&	10.05	&	7.52	&	7.69	\\  \hline
\multirow{2}{*}{VCTK-test (common)}	&	female		& 0.05 &		2.89	&	17.34	&	15.32	&	10.98	\\
	&	male 	& 0.05	&		{1.13}	&	9.89	&	8.19	&	4.80	\\ \hline
\multicolumn{3}{|c||}{\cellcolor{gray!7}Weighted average test} &		\cellcolor{gray!7}	3.79  & \cellcolor{gray!7}	11.81	& \cellcolor{gray!7} 9.18	& \cellcolor{gray!7} 7.77 \\
\Xhline{0.7pt}
  \end{tabular}  
\end{table*}

\begin{table}[h!]
  \caption{Primary utility evaluation: WER,\% achieved by $ASR_\text{eval}^\text{anon}$ on data processed by \textbf{B1.a}, \textbf{B1.b}, or \textbf{B2} vs.\ WER achieved by $ASR_\text{eval}$ on the original (Orig.) unprocessed data.}\label{tab:asr-results}
  \centering
  \renewcommand{\tabcolsep}{0.11cm}
 \begin{tabular}{|l||r||r|r|r|r|}
\Xhline{0.7pt}
 \multirow{2}{*}{\textbf{Dataset}}    & \multicolumn{4}{c|}{\textbf{WER,\%}}\\ \cline{2-5}
 &	\textbf{Orig.} & \textbf{B1.a} & \textbf{B1.b} & \textbf{B2~} \\ \hline \hline
LibriSpeech-dev	&		3.82	&	4.34	&	4.19	&	4.32	\\ \hline
VCTK-dev	&	10.79	&	11.54	&	10.98	&	11.76	\\ \hline
\cellcolor{gray!7}Average	dev &	\cellcolor{gray!7}  7.31  &  \cellcolor{gray!7}	7.94 & 	\cellcolor{gray!7} 	7.59 &  \cellcolor{gray!7} 8.04	\\ \hline \hline
LibriSpeech-test	&		4.15	&	4.75	&	4.43	&	4.58	\\ \hline
VCTK-test	&		12.82	&	11.82	&	10.69	&	13.48	\\ \hline
\cellcolor{gray!7}Average test &	\cellcolor{gray!7} 8.49 & \cellcolor{gray!7}		8.29 &	\cellcolor{gray!7} 7.56 & \cellcolor{gray!7} 9.03\\
\Xhline{0.7pt}
  \end{tabular}  
\end{table}
\normalsize

\begin{table}[h!]
  \caption{Secondary utility evaluation: 
  pitch correlation $\rho^{F_0}$ 
  and 
  gain of voice distinctiveness $G_{\text{VD}}$ achieved on data processed by \textbf{B1.a}, \textbf{B1.b}, or \textbf{B2}.}\label{tab:second-results}
  \centering
  \begin{tabular}{|l|l|c||r|r|r||r|r|r|}
\Xhline{0.7pt} 
\multirow{2}{*}{\textbf{Dataset}} & \multirow{2}{*}{\textbf{Gender}}  & \multirow{2}{*}{\textbf{Weight}} & \multicolumn{3}{c||}{ \makecell[c]{ \large$\boldsymbol{\rho}^{\boldsymbol{F}_0^{\textcolor{white}{'}}}$ } }   & \multicolumn{3}{c|}{ $\boldsymbol{G}_{\text{VD}}$} \\ \cline{4-9}
& & & \textbf{B1.a} & \textbf{B1.b} & \textbf{B2~}  & \textbf{B1.a} & \textbf{B1.b} & \textbf{B2~} \\ \hline \hline
\multirow{2}{*}{LibriSpeech-dev}	&	female & 0.25	&	0.77	&	0.84	&	0.64	&	-9.15	&	-4.92	&	-1.94	\\
	&	male  & 0.25	&	0.73	&	0.76	&	0.53	&	-8.94	&	-6.38	&	-1.65	\\ \hline
\multirow{2}{*}{VCTK-dev (different)}	&	female	 & 0.20 &	0.84	&	0.87	&	0.70	&	-8.82	&	-5.94	&	-1.32	\\
	&	male  & 0.20	&	0.78	&	0.76	&	0.59	&	-12.61	&	-9.38	&	-2.18	\\  \hline
\multirow{2}{*}{VCTK-dev (common)}	&	female  & 0.05	&	0.79	&	0.84	&	0.64	&	-7.56	&	-4.17	&	-1.14	\\
	&	male  & 0.05	&	0.72	&	0.72	&	0.54	&	-10.37	&	-6.99	&	-1.32	\\  \hline
\multicolumn{3}{|c||}{\cellcolor{gray!7}Weighted average dev}			& \cellcolor{gray!7} 0.77		&\cellcolor{gray!7}	0.80	&\cellcolor{gray!7}	0.61	&	\cellcolor{gray!7}	-9.71 &	\cellcolor{gray!7}	-6.44 &\cellcolor{gray!7}	-1.72	\\ \hline\hline
\multirow{2}{*}{LibriSpeech-test}	&	female & 0.25	&	0.77	&	0.85	&	0.61	&	-10.04	&	-5.00	&	-1.71	\\
	&	male & 0.25	&	0.69	&	0.72	&	0.54	&	-9.01	&	-6.64	&	-1.74	\\  \hline
\multirow{2}{*}{VCTK-test (different)}	&	female	& 0.20 &	0.84	&	0.87	&	0.68	&	-10.29	&	-6.09	&	-1.56	\\
	&	male & 0.20	&	0.79	&	0.77	&	0.66	&	-11.69	&	-8.64	&	-1.56	\\  \hline
\multirow{2}{*}{VCTK-test (common)}	&	female & 0.05	&	0.79	&	0.85	&	0.65	&	-9.31	&	-5.10	&	-1.59	\\
	&	male & 0.05	&	0.70	&	0.71	&	0.61	&	-10.43	&	-6.50	&	-1.36	\\ \hline
\multicolumn{3}{|c||}{\cellcolor{gray!7}Weighted average test}			& \cellcolor{gray!7} 0.77		&\cellcolor{gray!7}	0.80	&\cellcolor{gray!7}	0.62	&	\cellcolor{gray!7} -10.15	&	\cellcolor{gray!7} -6.44	&\cellcolor{gray!7}	-1.63	\\
\Xhline{0.7pt}
  \end{tabular}  
\end{table}

\subsection{Alternative anonymization systems}

In addition to the proposed three baseline systems described above, several alternative solutions have been developed by the
the challenge organizers and proposed as sources of additional inspiration for challenge participants.\footnote{Source code is available from  \url{https://github.com/Voice-Privacy-Challenge/Voice-Privacy-Challenge-2022}.}

\subsubsection{Alternative x-vector extractor}

The first alternative is based on the Sidekit toolkit~\cite{larcher2016extensible}.\footnote{The code is available in the \href{https://github.com/Voice-Privacy-Challenge/Voice-Privacy-Challenge-2022/tree/sidekit}{sidekit} branch.}
It replaces the  
Kaldi-based x-vector extractor with a Pytorch-based implementation.
This alternative system allows more straightforward modification of the extractor and adds more recent speaker verification loss functions, such as the additive angular margin loss \cite{Deng2021ArcFaceAA}.
X-vector representations are necessary to anonymize speech in the \textbf{B1} baselines. 
The provided x-vector extractor model is based on a ResNet-34 network and has an x-vector embedding size of 256.

Anonymization systems developed using the provided models will not be considered for evaluation and ranking in the challenge since they rely on additional data used for data 
augmentation\footnote{Room impulse response and noise database: \url{http://www.openslr.org/resources/28/}; MUSAN corpus of music, speech, and noise recordings: \url{http://www.openslr.org/resources/17/}.}
besides those specified in Section~\ref{sec:data}.  Nevertheless, participants may experiment with these  models  and report the results in their challenge papers.

\subsubsection{Alternative speech synthesis models}

The second alternative includes:\footnote{The code is available in the \href{https://github.com/Voice-Privacy-Challenge/Voice-Privacy-Challenge-2022}{master} branch and options for different speech synthesis models are setup by parameter  \texttt{tts\_type}.}
\begin{itemize}
    \item \texttt{am\_nsf\_pytorch}: A Pytorch-based re-implementation of \textbf{B1.a}. All the scripts and codes are unified under a Pytorch-based project. This is expected to be easier to customize and revise than the C++/CUDA-based \textbf{B1.a};
    \item \texttt{joint\_hifigan}: A variant of \textbf{B1.a}. The only difference is that the HiFi-GAN NSF model is replaced with the original HiFi-GAN. 
\end{itemize}

\subsubsection{Using self-supervised learning models}

The third alternative is based on self-supervised learning.\footnote{The code is available in the \href{https://github.com/Voice-Privacy-Challenge/Voice-Privacy-Challenge-2022/tree/ssl_anon_v2}{ssl\_anon\_v2} branch.}
Compared to \textbf{B1.b}, it has two main differences:
1) The BN features are replaced by the representation obtained from the last layer of a fine-tuned wav2vec 2.0 model \cite{baevski2020wav2vec}. Specifically, a wav2vec 2.0 Base model released by Facebook Research\footnote{https://github.com/facebookresearch/voxpopuli} was trained with 10k hours of unlabeled cross-lingual speech \cite{wang-etal-2021-voxpopuli} and finetuned on the labeled \textit{LibriSpeech-train-clean-100} data. The fine-tuning was conducted using the Fairseq toolkit{\footnote{\label{fairseq}{\url{https://github.com/pytorch/fairseq/}}}} with default settings;
2) HiFi-GAN is used as the speech waveform generation model.\footnote{The fine-tuned wav2vec 2.0 model and HiFi-GAN models are available at \url{https://zenodo.org/record/6350122/files/ssl_models.tar.gz}}
The other settings remain the same as in \textbf{B1.b}.

Anonymization systems developed using the provided 
SSL models will not be considered for evaluation and ranking in the challenge since they rely on additional data besides those specified in Section~\ref{sec:data}. Nevertheless, participants may experiment with these or other self-supervised models \cite{miao2022language,miao2022language2} and report the results in their challenge papers.

\section{Evaluation rules}

\begin{itemize}
    \item Participants are free to develop their own anonymization systems, using components of the baselines or not. 
     They are strongly encouraged to make multiple submissions corresponding to different EER thresholds (see Section~\ref{sec:asv-eval}).  Thresholds are applied to the weighted average of EER across  the VoicePrivacy test datasets (with weights of 0.5, 0.1 and 0.4  for \textit{LibriSpeech-test-clean}, \textit{VCTK-test (common)} and \textit{VCTK-test (different)}, respectively).
     \item The primary metrics (EER, WER) will be used for system ranking. Within each  interval for  EER~-- [15,20), [20,25), [25,30), [30,100)  -- systems  will be ranked in order of increasing    WER (averaged over \textit{LibriSpeech-test-clean} and \textit{VCTK-test}).
    All submissions considered for ranking should achieve a minimum average pitch correlation of $\rho^{F_0}>0.3$ for each development and test dataset.  
    \item  Participants can use only the training and development datasets specified in Section~\ref{sec:data} in order to train their system and tune hyperparameters. The use of any additional speech data is strictly prohibited.
    \item Participants must anonymize the development and test data in a \textit{speaker-level} manner. All enrollment (resp.\ trial) utterances from a given speaker must be converted into the same pseudo-speaker, and enrollment (resp.\ trial) utterances from different speakers must be converted into different pseudo-speakers. Also, the pseudo-speaker corresponding to a given speaker in the enrollment set must be different from the pseudo-speaker corresponding to that same speaker in the trial set.
    \item Participants must anonymize the dataset (\textit{LibriSpeech-train-clean-360}) used for training of  the evaluation models $ASV_\text{eval}^\text{anon}$ and $ASR_\text{eval}^\text{anon}$ using the same anonymization system applied to the development and test data, albeit in an \textit{utterance-level} manner. They must then train the evaluation models on the anonymized training data and apply them to the anonymized development and test data using the provided scripts. Modifications to the training or evaluation recipes (e.g., changing the network architecture, the hyperparameters, etc.) are prohibited.
    \item For every submitted system, participants must compute the primary objective metrics (WER, EER) and the secondary objective metrics ($\rho^{F_0}$, $G_{\text{VD}}$) for the two development datasets and the two test datasets using the provided evaluation scripts and retrained evaluation models. The organizers will be responsible for subjective evaluation only.

\end{itemize}

\section{Post-evaluation analysis}\label{sec:posteval}

The organizers will run additional post-evaluation experiments in order to further characterize the performance of submitted systems. To do so, we will ask all participants to share with us the anonymized speech data obtained when running their anonymization system on the training, development and test datasets. Further details will follow in due course.

\section{Registration and submission of results}

\subsection{General mailing list}

All participants and team members are encouraged to subscribe to the general mailing list.  Subscription can be done by sending an email to:

\begin{center} \href{mailto:sympa@lists.voiceprivacychallenge.org?subject=subscribe 2022}{sympa@lists.voiceprivacychallenge.org}
\end{center}

\noindent with \textit{`subscribe 2022'} as the subject line.  Successful subscriptions are confirmed by return email.  To post messages to the mailing list itself, emails should be addressed to:

\begin{center}
\href{mailto:2022@lists.voiceprivacychallenge.org?subject=voiceprivacychallenge 2022}{2022@lists.voiceprivacychallenge.org}
\end{center}

\subsection{Registration}

Participants/teams are requested to register for the evaluation.  Registration should be performed \textbf{once only} for each participating entity and by sending an email to:

\begin{center}
\href{mailto:organisers@lists.voiceprivacychallenge.org?subject=VoicePrivacy 2022 registration}{organisers@lists.voiceprivacychallenge.org}
\end{center}

\noindent with \textit{`VoicePrivacy 2022 registration'} as the subject line.  The mail body should include:
(i)~the name of the team; 
(ii)~the name of the contact person; 
(iii)~their affiliation;
(iv)~their country; 
(v)~their status (academic/non-academic).

\subsection{Submission of results}\label{subsec:submission}

Each participant may submit as many systems as they wish for each EER threshold provided in Section~\ref{sec:asv-eval}.  
In the case of multiple submissions for each condition, the organisers will use the single system with the lowest WER for ranking.  
Participants should submit audio data for only a single system per condition. Audio data for this system will be used for subjective evaluation.

Each single submission should include:

\begin{enumerate}
    \item The \textit{results} files generated by the evaluation scripts, which contain EER, WER, $\rho^{F_0}$ and $G_\text{VD}$ estimates for both development and test datasets,\footnote{Example \textit{results} files for the baseline system \textbf{B1.b}:
    \begin{itemize}
        \item
        Primary and secondary metrics: 
        \url{https://github.com/Voice-Privacy-Challenge/Voice-Privacy-Challenge-2022/blob/master/baseline/results/RESULTS_summary_tts_joint_nsf_hifigan} (saved in   \textcolor{darkspringgreen}{\texttt{exp\slash results-<date>-<time>}\slash results\_summary.txt})
        \item Additional metrics obtained using $ASV_\text{eval}$ and $ASR_\text{eval}$: \url{https://github.com/Voice-Privacy-Challenge/Voice-Privacy-Challenge-2022/blob/master/baseline/results/results.orig_tts_joint_nsf_hifigan} 
        (saved in   \textcolor{darkspringgreen}{\texttt{exp\slash results-<date>-<time>.orig}\slash results.txt})
        \item Additional metrics obtained using $ASV_\text{eval}^\text{anon}$ and $ASR_\text{eval}^\text{anon}$: \url{https://github.com/Voice-Privacy-Challenge/Voice-Privacy-Challenge-2022/blob/master/baseline/results/results.anon_tts_joint_nsf_hifigan} (saved in   \textcolor{darkspringgreen}{\texttt{exp\slash results-<date>-<time>}\slash results.txt})
    \end{itemize}
    } along with the full contents of the two \textit{results} directories \textcolor{darkspringgreen}{\texttt{exp\slash results-<date>-<time>}} and \textcolor{darkspringgreen}{\texttt{exp\slash results-<date>-<time>.orig}} also generated by the evaluation scripts;
    \item The corresponding PLDA (LLR) scores in Kaldi format (for the development and test data) obtained with the provided scripts;
    \item The corresponding anonymized speech  data (wav files, 16~kHz, with the same names as in the original corpus) generated from the development and test datasets. 
     For evaluation, the wav files will be converted to 16-bit signed integer PCM format, and this format is recommended for submission.
    These data will be used by the challenge organizers to verify the submitted scores, perform post-evaluation analysis with other metrics and subjective listening tests.
    All anonymized speech data should be submitted in the form of a single compressed archive.

\end{enumerate}

Each participant should also submit a single, detailed system description. 
All submissions should be made according to the schedule below. Submissions received after the deadline will be marked as `late' submissions, without exception.
System descriptions will be made publicly available on the Challenge website.
Further details concerning the submission procedure 
will be published via the participants mailing list and via the \href{https://www.voiceprivacychallenge.org/}{VoicePrivacy Challenge website}.

\section{VoicePrivacy Challenge workshop at INTERSPEECH 2022}

The VoicePrivacy 2022 Challenge will culminate in a joint workshop held in Incheon, Korea in conjunction with \href{http://www.interspeech2022.org/}{\textbf{INTERSPEECH 2022}} and in cooperation with the ISCA SPSC Symposium.\textsuperscript{\ref{fn:spsc}}
VoicePrivacy 2022 Challenge participants are encouraged to submit papers on the topic of their challenge entry according to the paper submission schedule
(see Section~\ref{sec:schedule}).
Paper submissions must conform to the format of the ISCA SPSC Symposium proceedings, detailed in the author’s kit\footnote{https://interspeech2022.org/files/IS2022\_paper\_kit.zip}, and be 4 to 6 pages long excluding references. Papers must be submitted via the online paper
submission system. 
Submitted papers will undergo peer review via the regular
ISCA SPSC Symposium review process, though the review criteria applied to regular papers will be adapted for VoicePrivacy Challenge papers to be more in keeping with systems descriptions and results.
Nonetheless, 
the submission of regular scientific papers related to voice privacy and anonymization are also invited and will be subject to the usual review criteria.
Since subjective evaluation results will be released only after the submission deadline, challenge papers should report only objective evaluation results.
The same paper template should be used for system descriptions but may be 2 to 6 pages in length.

Accepted papers will be
presented at the joint ISCA SPSC Symposium and VoicePrivacy Challenge Workshop and will be published as other symposium proceedings 
in the ISCA Archive. Challenge participants without accepted papers are also invited to participate in the workshop and present their challenge contributions
 reported in system descriptions. 

More details will be announced in due course.

\section{Schedule}\label{sec:schedule}

The result submission deadline is 31st July 2022.
All participants are invited to present their work at the joint SPSC Symposium and VoicePrivacy Challenge workshop that will be organized in conjunction with INTERSPEECH~2022.

\begin{table}[tbh]
  \caption{Important dates}\label{tab:dates}
  \centering
   \resizebox{\textwidth}{!}{
  \renewcommand{\tabcolsep}{-0.12cm} 
  \begin{tabular}{l r }
    \toprule
 Release of  training, development and evaluation data & \textcolor{blue}{Done} \\ \midrule
  Release of evaluation software and baselines & \textcolor{blue}{19th March 2022} \\ \midrule
Submission of challenge papers to the joint SPSC Symposium and VoicePrivacy Challenge workshop & \textcolor{blue}{15th June 2022} \\  \midrule
Author notification for challenge papers & \textcolor{blue}{1st July 2022} \\  \midrule 
Early bird registration to the joint SPSC Symposium and VoicePrivacy Challenge workshop &  \textcolor{blue}{7th July 2022} \\  \midrule
 Deadline for participants to submit objective evaluation results, anonymized data, and system descriptions & \textcolor{blue}{31st July 2022} \\  \midrule 
Final  paper upload   & \textcolor{blue}{5th September 2022} \\  \midrule
Joint SPSC Symposium and VoicePrivacy Challenge workshop   & \textcolor{blue}{23rd--24th September 2022} 
 \\ \bottomrule
   \end{tabular}
   }
\end{table}

\section{Acknowledgement}
This work was supported in part by the French National Research Agency under project
DEEP-PRIVACY (ANR-18-CE23-0018) and
jointly by the French National Research Agency and the Japan Science and Technology Agency under project VoicePersonae. 

\bibliographystyle{IEEEtran}
\bibliography{main}

\end{document}